\pdfoutput=1
\documentclass[conference]{IEEEtran}
\IEEEoverridecommandlockouts
\usepackage{cite}
\usepackage{amsmath,amssymb,amsfonts}
\usepackage{algorithmic}
\usepackage{graphicx}
\usepackage{textcomp}
\usepackage{xcolor}
\usepackage{array}
\usepackage{booktabs}
\usepackage{longtable}
\usepackage{tabularx}
\usepackage{tabulary}
\usepackage{blindtext, subfig}
\usepackage{dblfloatfix} % fix for bottom-placement of figure

\def\BibTeX{{\rm B\kern-.05em{\sc i\kern-.025em b}\kern-.08em
    T\kern-.1667em\lower.7ex\hbox{E}\kern-.125emX}}
\begin{document}

\title{Adapting a Container Infrastructure for Autonomous Vehicle Development}
\author{
\IEEEauthorblockN{Yujing Wang}
\IEEEauthorblockA{\textit{Department of Mechanical and Mechatronics Engineering} \\
\textit{University of Waterloo}\\
Waterloo, ON, Canada \\
yj9wang@edu.uwaterloo.ca} \and
\IEEEauthorblockN{Qinyang Bao}
\IEEEauthorblockA{\textit{Department of Mechanical and Mechatronics Engineering} \\
\textit{University of waterloo}\\
Waterloo On, Canada \\ 
q7bao@edu.uwaterloo.ca}
}

\maketitle

\begin{abstract}
In the field of Autonomous Vehicle (AV) development, having a robust yet flexible infrastructure enables code to be continuously integrated and deployed, which in turn accelerates the rapid prototyping process. The platform-agnostic and scalable container infrastructure, often exploited by developers in the cloud domain, presents a viable solution addressing this need in AV development. Developers use tools such as Docker to build containers and Kubernetes to setup container networks. This paper presents a container infrastructure strategy for AV development, discusses the scenarios in which this strategy is useful and performs an analysis on container boundary overhead, and its impact on a Mix Critical System (MCS). An experiment was conducted to compare both operation runtime and communication delay of running a Gaussian Seidel Algorithm with I/O in four different environments: native OS, new container, existing container, and nested container. The comparison reveals that running in containers indeed adds a delay to signal response time, but behaves more deterministically and that nested container does not stack up delays but makes the process less deterministic. With these concerns in mind, the developers may be more informed when setting up the container infrastructure, and take full advantage of the new infrastructure while avoiding some common pitfalls. 
\end{abstract}


\section{Introduction}
Agile, a new approach to software development, has been quickly winning favors with cloud developers over the traditional waterfall model. In agile practice, software developers write code, run it through CI/CD pipelines, integrate daily, and deploy as soon as a new feature needs to be tested \cite{b1}. This practice gives developers chances to test newly developed prototypes in all kinds of scenarios much more frequently. As a result, more iterations can be performed, and more bugs can be discovered in the meantime; thus, increasing both the quality and speed of the software. Applying agile practice to AV development is a bit more challenging; however, software technologies used on an AV often come from a wide range of temporal criticality: from low-level safety-critical mechanical controls to embedded realtime systems to high-level perception models, as well as mid-level networking applications. Fearing to mess up the temporal separation in an MCS, developers are generally hesitant to devise a unified infrastructure strategy. Furthermore, the business practice of modularizing teams into fine-grained functional unit also enhances the status quo mindset where things should be done as it is. As a result, each team must perform repetitive adaptation processes for each vehicle during each iteration. This repetitive procedure greatly slows down the development and testing feedback loop. 

\begin{figure}[htbp]
\centerline{\includegraphics[width=\linewidth,scale=0.5]{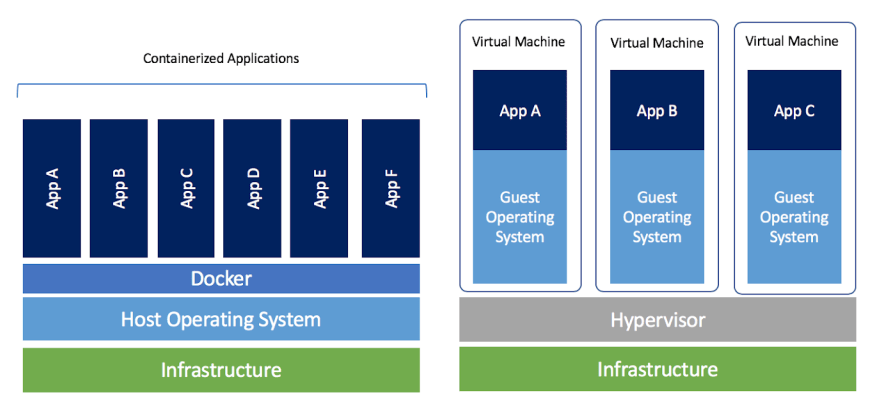}}
\caption{Containers vs Virtual Machine Infrastructure \cite{b4}}
\label{figure1}
\end{figure}

A well-designed container infrastructure removes these overheads and allows developers to build once and run on any platform. The accelerated build and test cycle makes continuous delivery of new features in response to ever-changing requirements possible \cite{b1}. The idea of a container first came from Linux namespace, which makes dedicating an exclusive resource set for a task possible \cite{b2}. Docker later came out to streamline the resource isolation process. Docker packages all software dependencies and running mechanism into an isolated environment called "image" Each software dependency or each step of the running mechanism is a layer in the image  \cite{b3}. To update the image, Docker updates the corresponding layer without making modifications to the rest of the image. Docker then deploys such an image into containers independent from other containers and the host environment. Docker daemon supplies the needed resource from the host machine to each container, thereby saving the developer from having to deploy an entire OS for each application, which is necessary for a virtual machine infrastructure, as shown in Fig.~\ref{figure1}. 

\begin{figure}[htbp]
\centerline{\includegraphics[width=\linewidth]{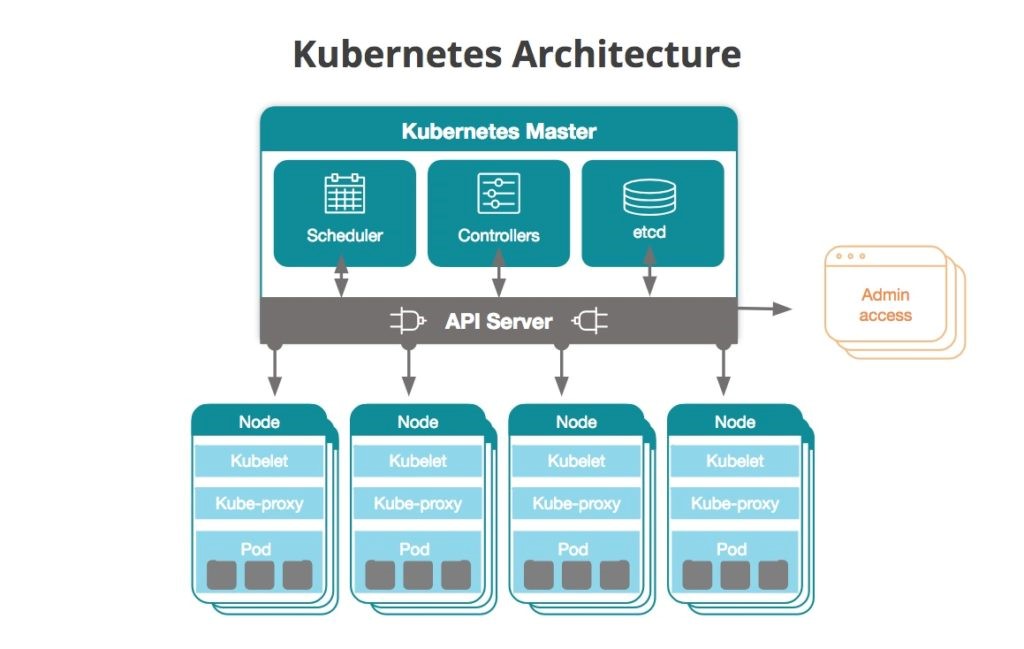}}
\caption{Architecture of Kubernetes\cite{b6}}
\label{figure2}
\end{figure}

As the number of containers increases, the need for establishing an efficient container network becomes crucial, especially for AV, where the storage and computing power is highly constrained. Kubernetes is an open source container orchestration tool, that lets the developer manage a network of containers \cite{b2}. Kubernetes reads declarative configurations from YAML files, in which the developer has specified the desired state. Knowing the current state and the desired state, Kubernetes works its way towards the desired state \cite{b5}. Kubernetes automates the tedious process of spawning, updating, and healing any number of containers. Moreover, it lets developers provision system resources for any containers. Fig.~\ref{figure2} depicts the master workers architecture of Kubernetes \cite{b6}. The master node accepts the command from the user, stores configuration, schedules pods, and realizes actions by sending signals to worker nodes. Worker nodes are connected to the master through Kube-proxy. Once a signal is received from the master node, Kubelet in each worker node executes the action accordingly. 

This paper will present the scenarios in which a container infrastructure benefits AV development, examine the resource and time overhead each container layer adds.

\section{Architecture and Applicable Scenarios}
\subsection{Multiple Variation for one Scenario}
\begin{figure}[htbp]
\centerline{\includegraphics[width=\linewidth,scale=0.5]{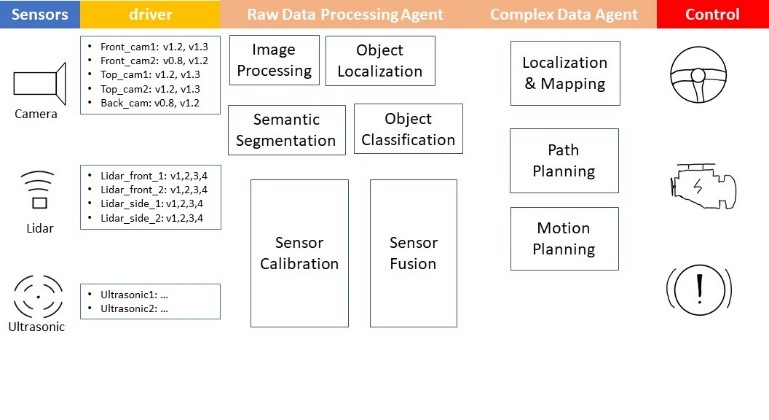}}
\caption{Line of Data Agents on a AV}
\label{figure3}
\end{figure}

In an AV, complex tasks such as lane changing, parking, and merging/yielding actions rely on a line of agents operating on data: data are collected, analyzed, and according to which actions are executed. Most actions are performed by local agents; some are performed by remote agents connected to each vehicle via the internet. By utilizing cloud computing, a vehicle can perform much more powerful data analysis that the limited local resource cannot support. This combination of local and remote agents mix makes up a Cyber Physical System (CPS) \cite{b7}. Fig.~\ref{figure3} shows the data processing line.

Located at the very beginning of the pipeline are the data collectors. These are sensors such as lidars and cameras. On top of each sensor is its respective driver. There may be multiple sensors of the same type mounted on the vehicle, that are highly similar but not made of the identical hardware. Take the on-vehicle camera as an example, the front\_cam1 optimized for traffic observation is different from the driver back\_cam adapted for rear approach checking. Having to manage different versions and variations of camera drivers is tedious, especially when one needs to perform A/B split testing to see which version performs better \cite{b1}. In a container infrastructure, the user packages each revision and versions in Docker images, then specify `sensor\_type:variation\_version`. For example, the developer may name the front left camera driver's version 1.2 as: `camera\_driver:frontLeft\_v1.1`. Using Kubernetes' replicable and self-healing deployment object, the developer can write a helm template manifest (via built-in Helm Templating Engine) for shelving the sensor driver containers \cite{b7}. Then specify the correlation between physical sensors and containers in a key-value file. 

Similarly, data processing agents down the data pipeline can be broken down in a similar structure using the aforementioned strategy. Each agent stands alone in one container. It receives input from upstream agents, processes it, and subsequently sends results to downstream agents. 

\subsection{One Module used in Multiple Scenarios}
The reverse is also true: A module can be packaged and employed in different scenarios. This enables application that handles one specific task to be repetitively deployed on different devices when a change in environment only affected its upstream or downstream agents. For example, running simulation is a very common practice to train vehicle’s perception module. The same perception module is coupled with different agents in the following scenarios (Fig.~\ref{figure4}) \cite{b8}: 
\begin{figure}[htbp]
\centerline{\includegraphics[width=\linewidth,scale=0.5]{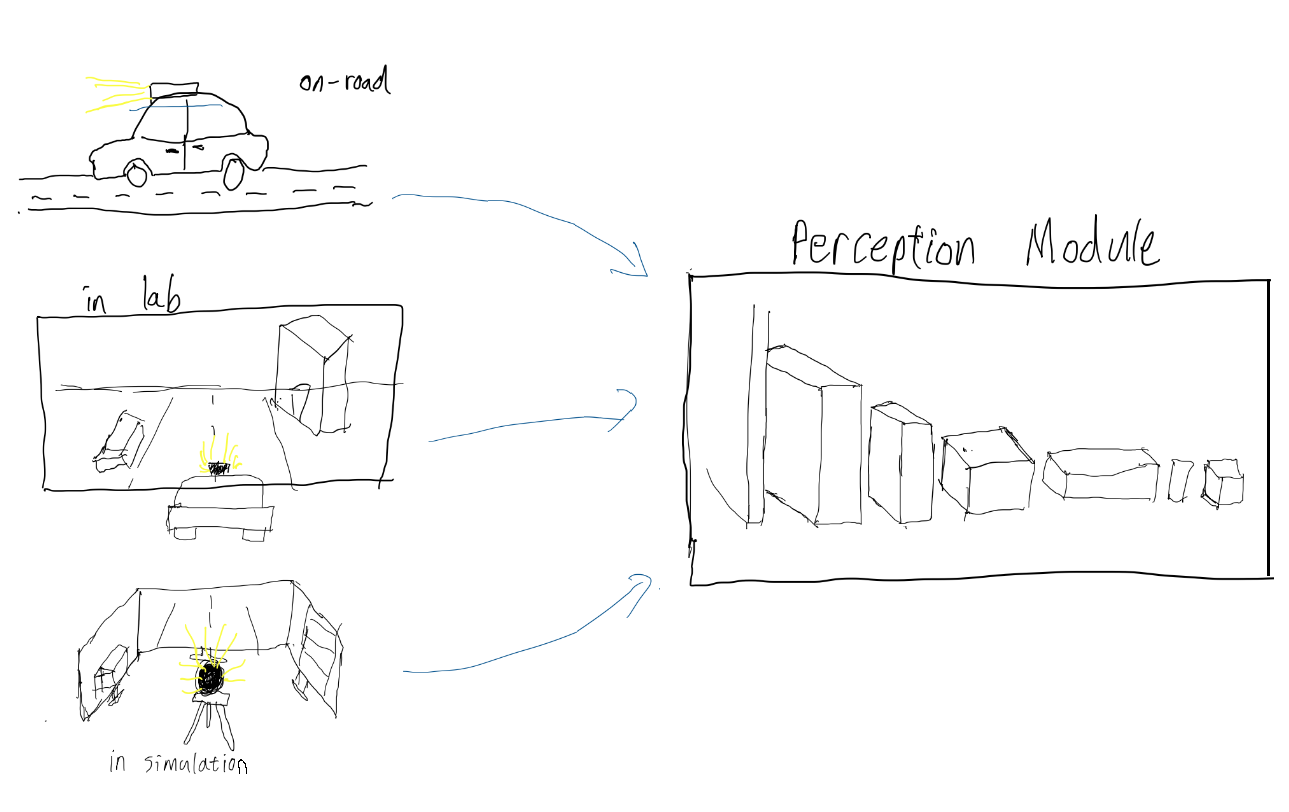}}
\caption{One Module Multiple Scenarios}
\label{figure4}
\end{figure}
\begin{enumerate}
    \item Running the vehicle on roads 
    \item Running the vehicle in a lab simulated environment
    \item Running perception core on a computer-simulated model.
\end{enumerate}
Although different setups are involved in different scenarios, the same perception module is used in all scenarios. To maintain consistency across all settings, the developers would deploy one perception model repetitively in all scenarios. 

\section{Resource Overhead and Container Boundary}
\subsection{Resource Overhead}
Overhead is a pain point when using containers. Running an application in a container inevitably consumes extra system resources and takes longer to communicate across the container boundary. In cloud computing, the limitation on memory and computation resource is close to negligible: developers can add more machines to the container network, thereby scaling the cluster horizontally to accommodate for increased usage. In AV development, the physical space on the vehicle for housing machines is limited. Fog Architecture proposes a way to utilize cloud computing's power best and accommodate for the limited space on a vehicle by having vehicles upstreaming resource intensive computation logic to edge devices \cite{b9}. Modules on these devices enable vehicles to navigate through more complex situations such as driving in a chaotic urban environment where pedestrians and vehicles may cross paths at irregular intervals and random locations. The vehicle itself, on the other hand, hosts a complete ecosystem of data processing agents to navigate through places where network connections are weak to non-existent, and the traffic is more predictable, such as driving on a highway in the countryside. The architecture of such infrastructure is similar to "One module used in multiple scenarios". Each vehicle joins the container network as a less powerful node. Each module is deployed repetitively in each vehicle node and cloud node. The container flavor manager manages which tag of the image will be deployed, given the specs of the target node. Though not the focus of this paper, containers infrastructure allows AVs to tap into the computing power of cloud machines and henceforth to circumvent the limitation on physical space constraints partially. 

\subsection{Overhead Analysis and Realtime Scheduling Analysis}
Even more limited than the resource is the response time in a time-critical system. A signal traversing in container networks needs first to exit its originating container and enter its destination container, crossing at least two layers of delay per container involved. The more containers involved, the more layers that the signal needs to cross, and the higher the delay stacks. The accumulation of delay worsens when there are containers nested inside of other containers or when signals are traversing through multiple intermediate containers. Compounding layers of boundary communication gives a significant delay in the signal relay process. The delay adds more considerable uncertainty in communication time, making the realtime system less deterministic. \cite{b10} performed a runtime analysis on the temporal criticality of each container's operation runtime in 4 different environments: ubuntu vanilla-native, vanilla-docker, RT-native, and RT-docker. The result showed that runtime in Docker is approximately the same as running natively. Real-time enabled Linux kernel performs more deterministically than the vanilla kernel, which nevertheless, is overall faster in both empty and loaded context. \cite{b10} did not study the effect of crossing the container itself, which we intend to investigate in the following experiment to understand how container network should be orchestrated to provide the lowest average delay and lowest uncertainty in the signal relay process. 

\subsection{Experimental Setup}
To study the communication delay across the container boundary, we decided to perform an experiment to see how container boundary affects communication time. Four scenarios are tested and juxtaposed: 
\begin{enumerate}
    \item Running on a native machine
    \item Running by spawning new containers
    \item Running in an existing container
    \item running in a five-folds nested container
\end{enumerate}

In this experiment, we use an algorithm that perform iterative Gaussian Seidel Approximation on a strictly diagonally dominant matrix \cite{b11}. For any Matrix operation in the form
\begin{equation}
A\bold{x}=b,\label{eq1}
\end{equation}
where $\bold{x}$ is an unknown $n \times 1$ vector, A is a known $n \times n$ strictly diagonal matrix, and b is a known $n \times 1$ vector. Decomposing Matrix A into lower triangular matrix $L*$ and strictly upper part $U$ such that $(L*)+U = A$, we get
\begin{equation}
(L*)x=U-bx\label{eq2}
\end{equation}
Isolating for one x and using forward subsitution, 
\begin{equation}
x^{k+1}=(L*)^{-1} (b-Ux^k).\label{eq3}
\end{equation}
This gives us an iterative algorithm to obtain the next guess of each element $x_{i}^{k+1}$ from $x_{i}^{k}$ its previous guess, $x_{i}^{0}$ a base case "initial guess" using formula
\begin{equation}
x_{i}^{k+1}=\dfrac{1}{a_{ii}} (b_{i}-\sum_{j=1}^{i-1}a_{ij}x_{j}^{k+1}-\sum_{j=i+1}^{n}a_{ij}x_{j}^{k}), i =1,2,...n.
\label{eq4}
\end{equation}
This operation performs an iterative solution to calculate result. This effectively converts ordinary matrix solution with uncertain runtime to O(n) runtime during each iteration and storing O(n) variables in memory space. Knowing the size of the matrix, we have a constant runtime and constant memory, which are convenient for measurement purpose. Using an initial guess of [0,0,0,0,0], the program performs a total of 100 calls per experimental scenario and display the generated logs when finished. For each call, it reads the "initial guess" and signal sent time as input, records the signal relay time, performs 2500 iterations of the algorithm, writes results as logs to a local persistent storage, and then sends output back, which is used as the initial guess in the next call. The value used for matrix A and vector b, and the calculated solution for x are
\begin{equation}
A=\begin{bmatrix}
4&1&2&1&1\\3&5&1&1&1\\1&1&3&1&1\\1&1&1&5&1\\1&1&1&1&9
\end{bmatrix},
  b=\begin{bmatrix} 4\\7\\3\\9\\2 \end{bmatrix},\quad x=
 \begin{bmatrix} {39}/{106} \\ {46}/{53} \\ {11}/{106} \\ {329}/{212} \\ -{21}/{212} \end{bmatrix}.
 \label{eq5}
\end{equation}

Only output that falls within a 99.95\% confidence interval in a $5^{th}$ degree of freedom T-test will be used in the overhead analysis. The t value is determined using (\ref{eq6}), where $\Bar{x}$ is the answer being sampled, $\mu$ the mean is the “ground truth” calculated in II, n is the sample size, and $\hat{\sigma}$ is standard deviation of sample results \cite{b12}.
\begin{equation}
t=\frac{\Bar{x}-\mu}{\hat{\sigma} \/\sqrt{n}} \label{eq6}
\end{equation}

Note that since each call performs 2500 iterations, it is very unlikely to have the result not fall within the 99.95\% confidence level and that since the result of the previous stage is feed into the next step, the accuracy is likely to increase over time. 

In "running on native machine" and "running by spawning new containers", the function is exposed through "\_\_main\_\_.py". When "python gausse.py" is executed, an instance is initiated on the local machine or inside a new container. As soon as the results are returned, the instance is terminated, and its allocated memory space released. Comparing running code on the native machine against running code by spawning a new container each time help us study the overhead of instantiating a new container. In "running in an existing container" and "running in five-fold nested container in container", the function is wrapped in Flask, a python web framework that allows communication via HTTP calls. The application is only initiated once at the start, so we can measure the time for signals to traverse through the communication layer and compare how nesting containers affect the signal relay time. To keep the environment as consistent as possible, we performed all four scenarios on one machine with the following specification (note specs not related to this experiment are not listed):
\begin{itemize}
    \item CPU: Intel(R) Core(TM) i7-6700 CPU @ 3.40GHz
    \item Memory: 16 GiB Memory DDR3 at 1600MHz
    \item Storage: 2 TB HDD TOSHIBA DT01ACA200  
    \item File System: ext4 
    \item OS: Ubuntu 18.04.3 LTS
    \item Container: Docker v19.3
    \item Code base: Python v3.4
\end{itemize}
When the experiment is finished, the average and standard deviation of operation time and communication delay will be determined using (\ref{eq7}) and (\ref{eq8}), where n is the number of elements,  $\mu$ is average and $\sigma$ is standard deviation: 
\begin{equation}
    \mu=\dfrac{1}{n}\sum_{i=1}^{n}x_i \label{eq7}
\end{equation}

\begin{equation}
    \sigma=\sqrt{\dfrac{1}{n}\sum_{i=1}^{n}(x_i-\mu)^2}\label{eq8}
\end{equation}

\section{Results and Finding}

\begin{table*}[htbp]
\caption{Table Type Styles}
\begin{center}
\begin{tabular}{|c|c|c|c|c|c|}
\hline
\textbf{Measurement} & Time & Native & New Container & Existing Container & 5x Nested Containers  \\
\hline
Operation & $\Bar{t}$ & 
0.009868 &
0.022465 &
0.033195 &
0.030283465
 \\ 
\cline{2-6}
Duration & $\hat{\sigma}$ & 
0.000867726 &
0.001595 &
1.62101E-05 &
0.003111 
\\ 
\hline
Communication & $\Bar{t}$ & 
4.01E-07 &
2.440119 &
0.001893 &
0.001209655
\\ 
\cline{2-6}
Duration & $\hat{\sigma}$ & 
1.51E-07 &
0.45777 &
7.87E-07 &
0.000561769 
  \\ 
\hline
Accuracy & - & 100\% & 100\%  &  100\% &  100\% \\ 
\hline
\end{tabular}
\label{tab1}
\end{center}
\end{table*}

\begin{figure*}[htbp]
\centering
\subfloat[][]{
    \includegraphics[width=0.5\linewidth,scale=1]{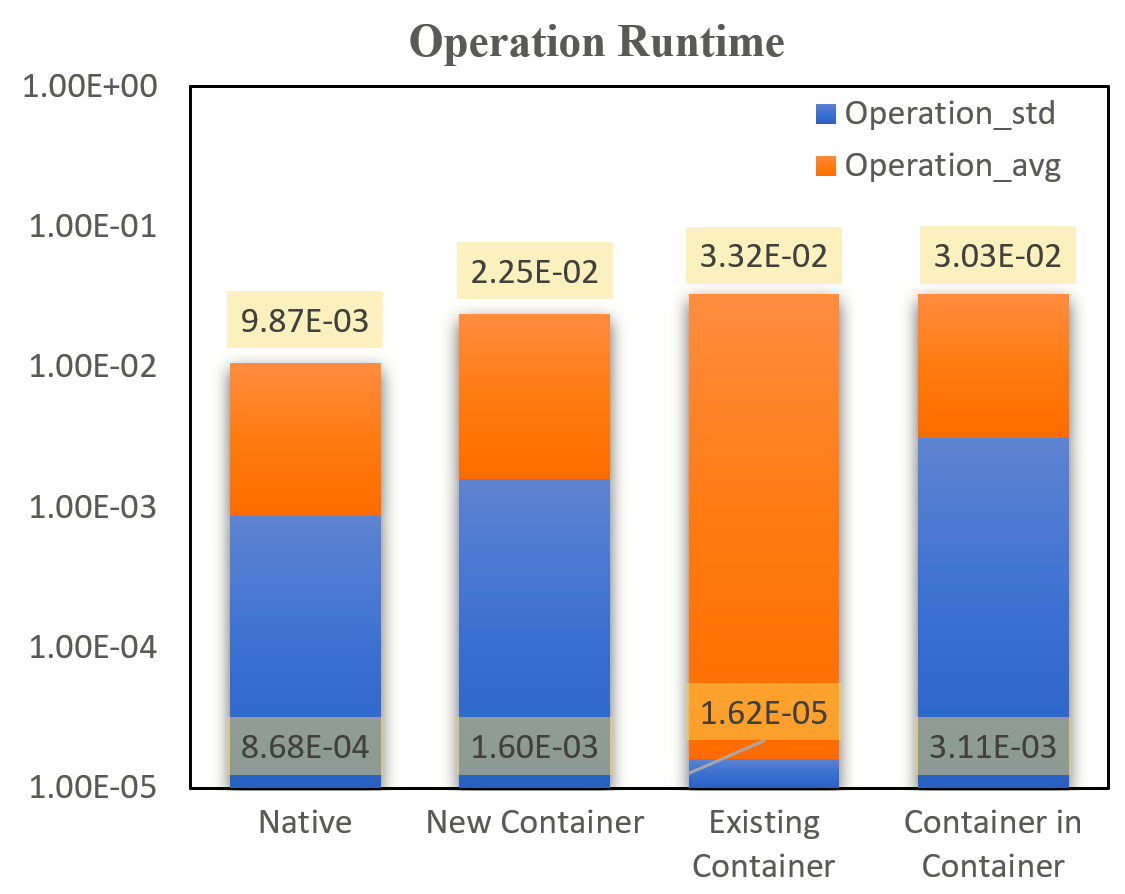}
}
\subfloat[][]{
   \includegraphics[width=0.5\linewidth,scale=1]{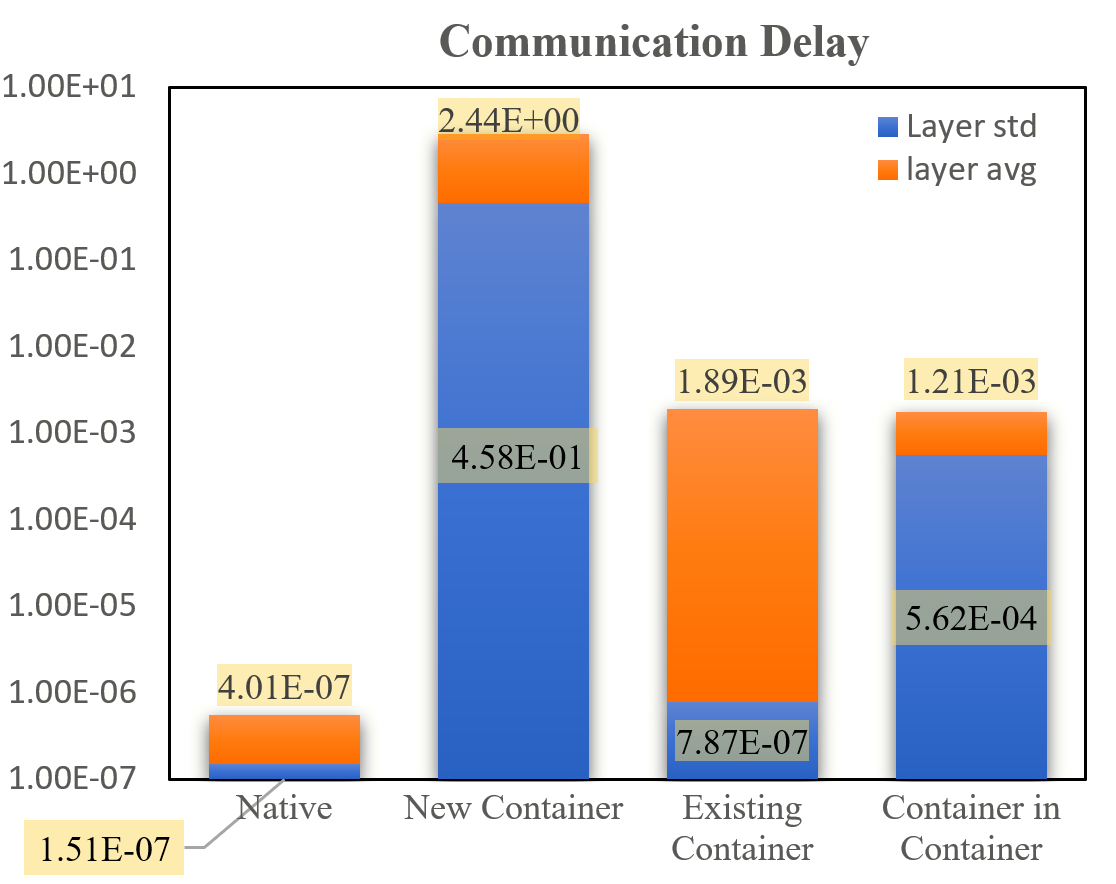}
}
\caption{ Runtime Average and Standard Deviation Comparison Graph of (a) Operation and (b) Layering }\label{figure5}
\end{figure*}

The logs and calculated data is hosted on GitHub repository kamagawa/containers\_infrastructure \cite{b13}, the average and standard deviation of operation and communication time are documented in Table I. From an accuracy perspective, all entries meet the accuracy requirement, thanks to strict diagonality of the matrix that guarantees convergence within 2500 iterations. Fig.~\ref{figure5}a shows that executing the algorithm on the native OS is much faster than any other option, then followed by running in new containers, nested containers, and existing containers. The standard deviation of running in an existing container is the lowest by a large margin, making it the most deterministic option. Having been provisioned its resource bundle, processes running inside Docker enjoy isolation from system noises that affect native processes. However, when placing containers inside containers five-folds, the docker daemon's scheduler can no longer provision the containers directly, thus giving rise to the standard deviation of container communication delay. Operation runtime in a five-folds nested container is approximately the same as running in a one-layer container. 

Fig.~\ref{figure5}b shows that the communication delay of the native process is the lowest by a large margin. There is no delay between the signal sent and received. Running code by spawning a new container is the longest and the least deterministic option since the process of creating a new container and provisioning it resource takes a long time. Sending signals across container boundary to existing containers and nested containers in containers consumes a roughly equal amount of time. However, sending signals to 1 layer of container is much more deterministic than five layers of nested containers. 

\begin{figure}
\centerline{\includegraphics[width=\linewidth,scale=0.5]{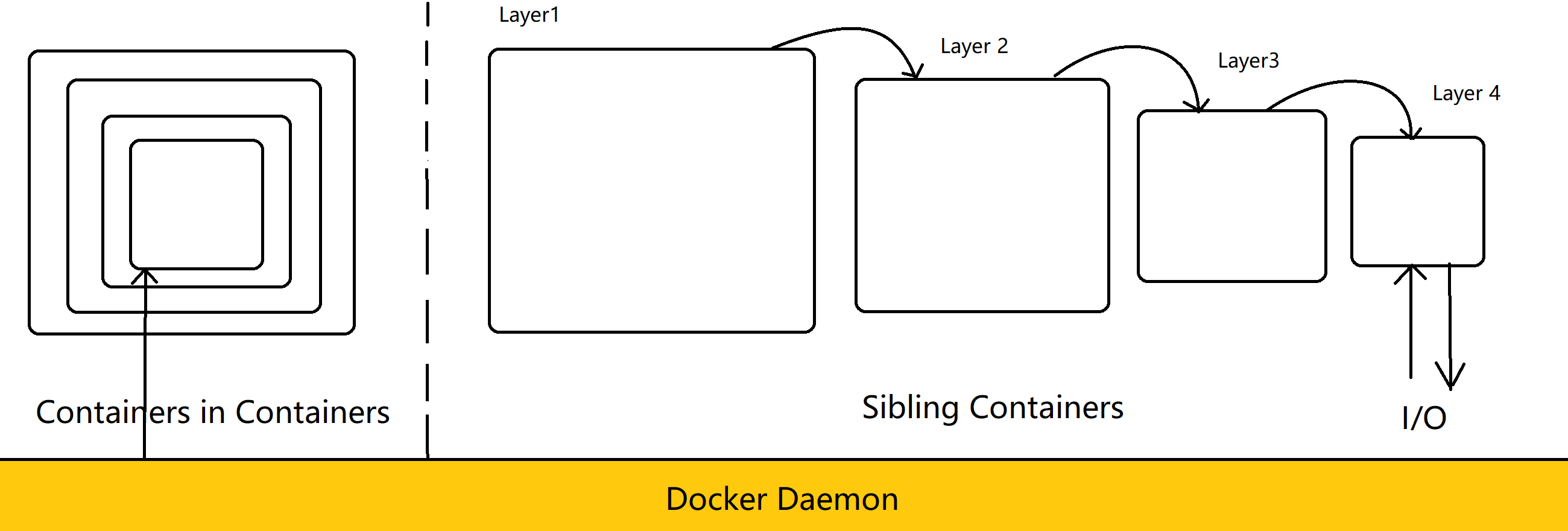}}
\caption{Containers vs Virtual Machine Infrastructure \cite{b4}}
\label{figure6}
\end{figure}

Whether it's operation time or communication time, a five-fold nested container doesn't take a longer time than a regular container. This finding is a little bit surprising, as one would expect crossing five layers of the container boundary would take five times as long as crossing one layer of container boundary. This phenomenon could be attributed to the architecture of container creation in Docker, as shown in Fig.~\ref{figure6}. When creating a new container inside an existing container, Docker Damon creates a sibling container that is linked to the current container, rather than directly spawning the new container inside the current one \cite{b14}. However, sibling containers' resource sets are provision from the current container's resource set. The weak resource isolation among sibling containers makes nested containers behave less deterministic than standard ones.  

Understanding the runtime behavior of containers in different situations is crucial to setting up a robust and flexible container infrastructure for Autonomous Vehicle development. For such a safety-critical system, being able to know when a task will fire, and finish is more important than finishing it as quickly as possible. When one needs to nest a container, they must ask themselves, is this task time-critical, and is it ok for it to share the resource with its current containers?  To achieve a higher level of temporal precision, industry partners often use a realtime enabled kernel (rt-kernel) and implement their Network Time Protocol server as part of their container networks.

\section{Conclusion and Future Work}

This paper presented a container infrastructure for autonomous vehicle development, the scenarios in which it outperforms native development such as "Multiple Variation for one Scenario" and "One Module used in Multiple Scenarios". We presented an option to extend the container network with the Fog network to tap into the power of cloud computing such that more powerful computation can be performed on cloud and less powerful computation perform locally. Then study the runtime overhead when running in container network compared to the native machine and discovered that signals crossing into and out of containers experience a considerable delay compared to native but behaves much more deterministically. Nested containers do not add extra overhead because architecturally, they are linked as "sibling containers" rather than being put inside one another. However, their provisioned resources are shared, making nested containers less deterministic than a single layer container. Understanding the operation runtime and communication delay is essential when designing a container network of a mixed critical system. To fulfill a higher degree of temporal precision, companies often use rt-kernel and implement their customized time control logic.

During the experiment, we face many obstacles that could potentially be an inspiration for future work. One is streamlining the process of image creation such that when a non-critical line is changed in the code, it doesn't rebuild the entire layer. Building on top of the container architecture, we will study a pragmatic approach for utilizing Kubernetes to orchestrate a robust network to run simulations and allows for  CI/CD of new features in AV development.

\end{document}